\documentclass[reprint,prb,superscriptaddress,notitlepage,longbibliography]{revtex4-1}

\usepackage{graphics,epsfig,amsfonts,amssymb,amsmath,color}
\usepackage[normalem]{ulem}
\usepackage{bm,bbm}
\usepackage[mathcal]{euscript}
\usepackage{color}
\usepackage{float}
\newcommand{\bra}[1]{\ensuremath{\langle#1|}}
\newcommand{\ket}[1]{\ensuremath{|{#1}\rangle}}

\newcommand{\beginsupplement}{%
		\setcounter{equation}{0}
		\renewcommand{\theequation}{S\arabic{equation}}
        \setcounter{table}{0}
        \renewcommand{\thetable}{S\arabic{table}}%
        \setcounter{figure}{0}
        \renewcommand{\thefigure}{S\arabic{figure}}%
        \renewcommand{\thesection}{S\arabic{section}}
     }

\begin{document}

\title{Enhancing the dipolar coupling of a $S$-$T_0$ qubit with a transverse sweet spot}
\author{J. C. Abadillo-Uriel}
\author{M. A. Eriksson}
\affiliation{Department of Physics, University of Wisconsin-Madison, Madison, WI 53706, United States}
\author{S. N. Coppersmith}
\affiliation{Department of Physics, University of Wisconsin-Madison, Madison, WI 53706, United States}
\affiliation{School of Physics, The University of New South Wales, Sydney 2052, Australia}
\author{Mark Friesen}
\affiliation{Department of Physics, University of Wisconsin-Madison, Madison, WI 53706, United States}

\begin{abstract}
A fundamental design challenge for quantum dot spin qubits is to extend the strength and range of qubit interactions while suppressing their coupling to the environment, since both effects have electrical origins.
Key tools include the ability to retune the qubits, to take advantage of physical resources in different operating regimes, and to access optimal working points, or ``sweet spots," where dephasing is minimized.
Here, we explore an important, new resource for singlet-triplet qubits: a transverse sweet spot (TSS) that enables (i) direct transitions between qubit states, (ii) a strong, charge-like qubit coupling, and (iii) leading-order protection from electrical fluctuations.
Of particular interest is the possibility of transitioning between the TSS and symmetric operating points while remaining continuously protected. 
This arrangement is ideal for coupling qubits to a microwave cavity, because it combines maximal tunability of the coupling strength with leading-order noise suppression.
We perform simulations with $1/f$-type electrical noise, demonstrating that two-qubit gates mediated by a resonator can achieve fidelities $>99$\% under realistic conditions.
These results greatly expand the toolbox for singlet-triplet qubits.
\end{abstract}

\maketitle

Recent advances in semiconducting spin qubits \citep{LossPRA1998, KaneNature1998} have enabled single-qubit gates with high fidelities \citep{VeldhorstNatNano2014, Kawakami11738, KimNPJQI2015, TakedaSciAdv2016, YonedaNatNano2018}, and two-qubit exchange-based gates\citep{VeldhorstNature2015, Zajac439, WatsonNature2018, NicholNPJQI2017, XuePRX2019} with fidelities $>$94$\%$ \citep{DzurakArxiv}. 
While these exchange gates are relatively fast, their interaction range is limited -- typically to nearest neighbors.  
One method for increasing the interaction range is to insert an intermediary coupler, such as a superconducting microwave cavity \citep{RaimondRMP2001, BlaisPRA2004, ChildressPRA2004, SillanpaNature2007, MajerNature2007, XiangRMP2013, DeshuiQST2017}. 
However, strong qubit-resonator couplings have been difficult to realize, due to the small magnetic dipole of the spins\citep{ImamogluPRL2009, AmsussPRL2011, EichlerPRL2017}, which results in slow qubit gates. 
A common strategy for enhancing this coupling involves hybridizing the spin and charge degrees of freedom via the spin-orbit interaction, which arises naturally in GaAs, and can be induced by micromagnets in Si \citep{XuedongPRB2012, BenitoPRB2017}. 
In this way, strong coupling has been achieved in both GaAs and Si \citep{LandigNature2018, MiNature2018, Samkharadzeeaar4054}. 
However the gates are still slow and susceptible to electrical (``charge") noise, motivating a search for new methods to enhance the qubit charge dipole, as well as sweet spots to suppress the effects of noise.

\begin{figure}[t]
\includegraphics[width=1.8 in]{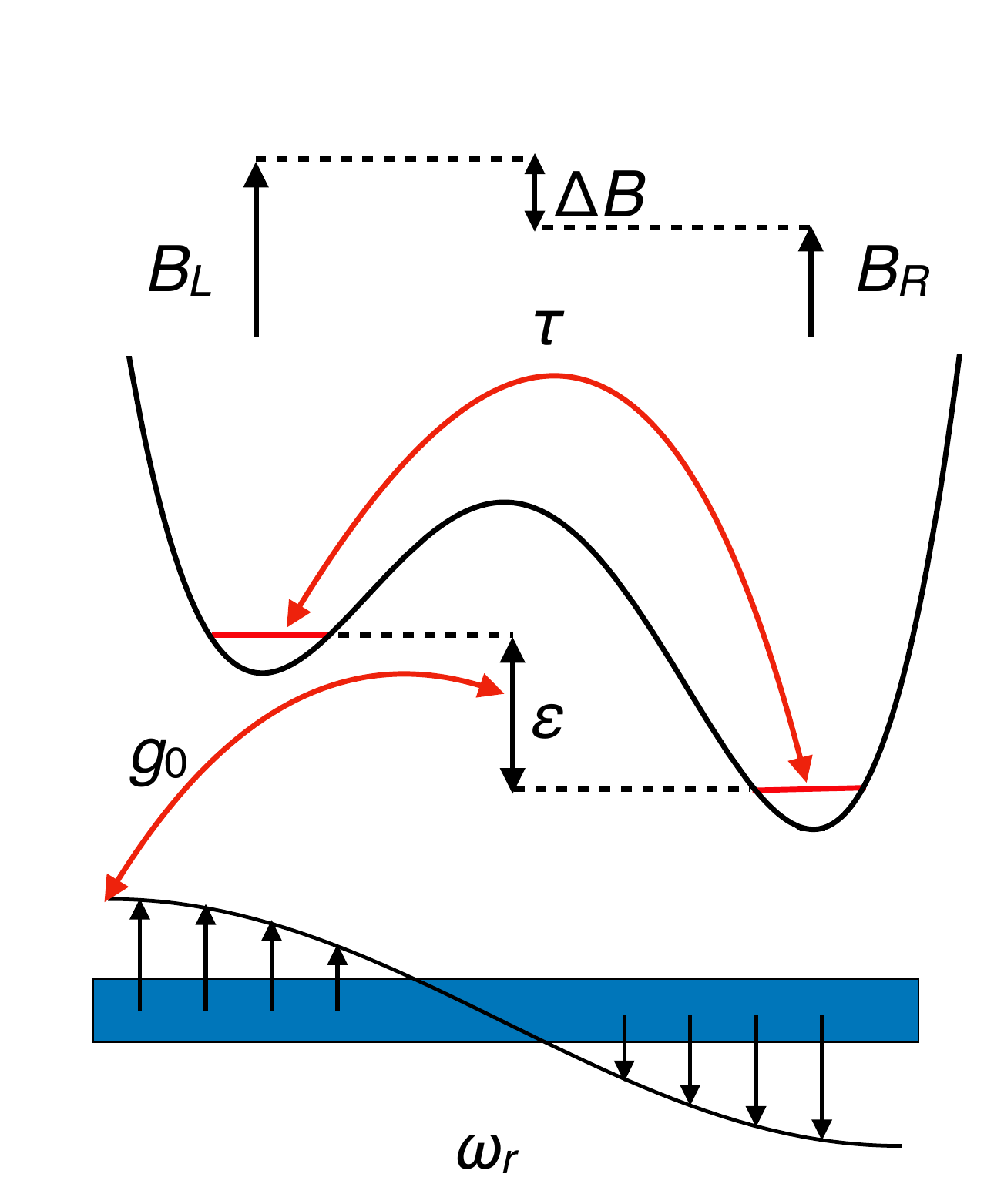}
\caption{
Double-dot device schematic, including an optional coupling to a superconducting microwave resonator.
A singlet-triplet qubit is formed in a double dot containing two electrons.
$\varepsilon$ is the detuning between the two sides of the device, 
$\tau$ is the corresponding tunnel coupling, $\Delta B$ is the Zeeman energy associated with the inter-dot magnetic field difference (or ``gradient"), and
$g_0$ is the optional capacitive coupling between a qubit plunger gate and the anti-node of a resonator of frequency 
$\omega_r$, which can be used to mediate two-qubit gate operations.}
\label{fig1}
\end{figure} 

\begin{figure*}[th]
\includegraphics[width=7.3 in]{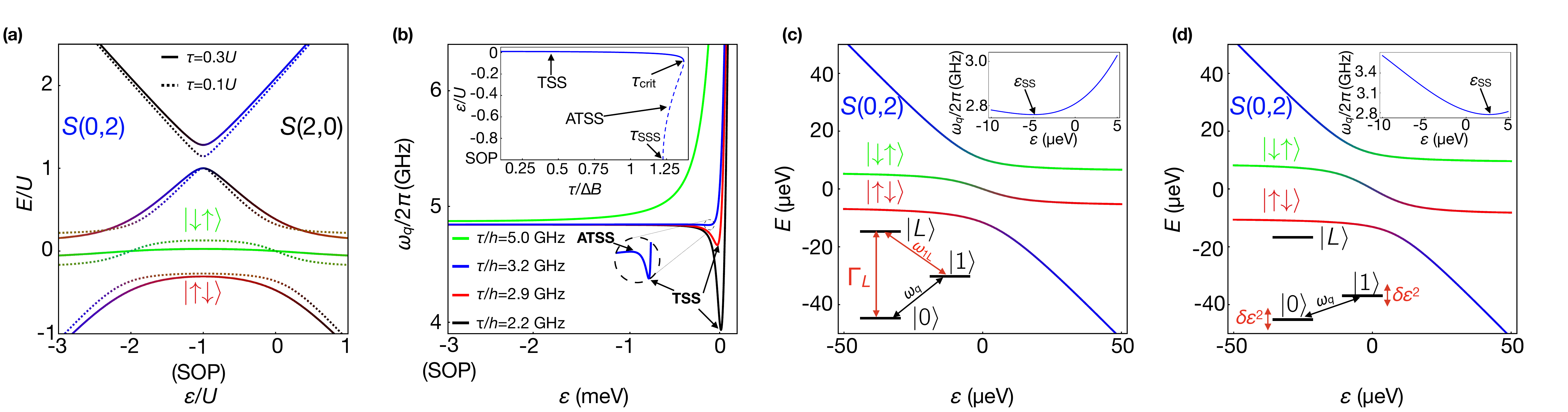}
\caption{Operating regimes of the transversely coupled sweet spot (TSS) of a $S$-$T_0$ qubit. 
\textbf{a} Energy level diagrams obtained from Hamiltonian (1) assuming that the tunnel coupling $\tau$ and the field difference between the dots $\Delta B$ obey $\tau>\Delta B$ (solid lines), or $\tau<\Delta B$ (dashed lines). 
In both cases we take $\Delta B=0.2U$, where $U$ is the double-occupation charging energy. 
Singlet-triplet qubits are often operated at either the symmetric operating point (SOP), where $\varepsilon=-U$, or near the $S(1,1)$-$S(0,2)$ charging transition for the singlet state, where $\varepsilon=0$. 
\textbf{b}  
The TSS and  the alternative TSS are indicated on the qubit dispersions $\omega_q$, obtained for $\Delta B/h=2.5$ GHz and several different values of $\tau$.
We also assume $U=3$~meV here, and throughout this work.
For $\tau\leq \tau_{\rm crit}$, a TSS dip forms near $\varepsilon=0$ (black arrows). 
For $\tau_\text{SSS}\leq \tau\leq \tau_{\rm crit}$, a very shallow ATSS peak also emerges, to the left of the TSS (lower inset).
The values of $\tau_\text{crit}$ and $\tau_\text{SSS}$ both depend on $\Delta B$.
Upper inset: the location of the various sweet spots and critical points in detuning space, for $\Delta B/h=2.5$~GHz. 
\textbf{c, d} Energy level diagrams near the charging transition, showing two types of behavior (main panels), and their corresponding qubit energy splittings (upper insets) and sweet spots ($\varepsilon_\text{SS}$).
\textbf{c} For $\tau >\Delta B$, we observe $\varepsilon_\text{SS}<0$.
Degenerate energy levels ($\omega_q\approx\omega_{1L}$) near the TSS can induce unwanted excitations to the leakage state $\ket{L}$ (lower inset), with decay rate $\Gamma_L=1/T_L$. 
(Here, $\tau/h=1.75$~GHz and $\Delta B/h=1.5$~GHz.)
\textbf{d} For $\tau\lesssim\Delta B$, we observe $\varepsilon_\text{SS}>0$.
In this case, the excitation energies are no longer degenerate at the TSS, suppressing leakage, but the qubit is more charge-like, and therefore susceptible to fluctuations of the detuning parameter $\delta\varepsilon$.
At the sweet spot, the dominant fluctuations occur at $\mathcal{O}[\delta\varepsilon^2]$, causing weak dephasing.
(Here, $\tau/h=1.5$~GHz and $\Delta B/h=2.5$~GHz.)}
\label{fig2}
\end{figure*}

For singlet-triplet spin qubits\citep{Levy2002,PettaScience2005, wu2014two}, recent attention has focused on a sweet spot known as the symmetric operating point (SOP), due to its favorable coherence properties\citep{ReedSOP,Martins:2016}.
The position of the SOP -- as far as possible from the (2,0)-(1,1) or (1,1)-(0,2) charging transitions -- reduces its sensitivity to charge noise, but also suppresses its charge dipole moment. 
In this regime, the weak dipole coupling is mainly longitudinal in form\citep{Ruskov2017,HarveyPRB2018}, enabling $Z$ rotations and two-qubit CPHASE gates.
In contrast, the charge dipole increases near a charging transition -- particularly its transverse component, enabling $X$ rotations and two-qubit iSWAP gates. 
In this regime, when the inter-dot Zeeman energy difference (or ``gradient") $\Delta B=g\mu_B(B_L-B_R)$ is larger than the tunnel coupling, the transverse coupling can dominate over the longitudinal coupling\citep{GuidoPRB2006, Taylorarxiv, JinPRL2012}; however, the qubit also becomes more sensitive to charge noise. 

In this work, we investigate a family of sweet spots, with strong transverse couplings, located far from the SOP. 
We show that these transversely coupled sweet spots (TSS) represent an important new working regime for singlet-triplet qubits that can be exploited to perform high-fidelity single-qubit gates with AC electrical driving fields, or to enable capacitively coupled two-qubit gates. 
(Here, we focus on two-qubit gates mediated by a superconducting cavity.) 
We describe protocols for one and two-qubit gate operations that provide constant noise protection, even while transitioning between operating points. 
This allows us to take advantage of the resources available in different working regimes, and greatly enhances the toolbox for operating singlet-triplet qubits.

\vspace{.1in} \noindent 
\textbf{RESULTS} 
\vspace{.05in} \\ \noindent 
\textbf{TSS and SOP Sweet Spots.}
We initially assume that the global magnetic field $B$ is large enough that the polarized triplet states may be ignored. We include the polarized triplets later; however, the simpler model serves to illustrate the key physics.
In this case, the Hubbard Hamiltonian of a singlet-triplet qubit in the basis $\{|S(1,1)\rangle,|T_0(1,1)\rangle,|S(0,2)\rangle,|S(2,0)\rangle\}$ is given by\citep{GuidoPRB1999}
\begin{equation}
H_\text{ST}=\begin{pmatrix}
0 & \Delta B & \sqrt{2}\tau & \sqrt{2}\tau \\
\Delta B & 0 & 0 & 0 \\
\sqrt{2}\tau & 0 & -\varepsilon & 0 \\
\sqrt{2}\tau & 0 & 0 & 2U+\varepsilon
\end{pmatrix},
\end{equation}
where $\tau$ is the tunnel coupling between the two sides of the double dot (Fig.~1), $\varepsilon$ is the detuning between dots, and $U$ is the charging energy for doubly occupied states. 
Here we define $\varepsilon=0$ as the position of the $\ket{S(1,1)}$-$\ket{S(0,2)}$ charging transition, with the SOP located at $\varepsilon=-U$.

\begin{figure*}[t]
\includegraphics[width=6.5 in]{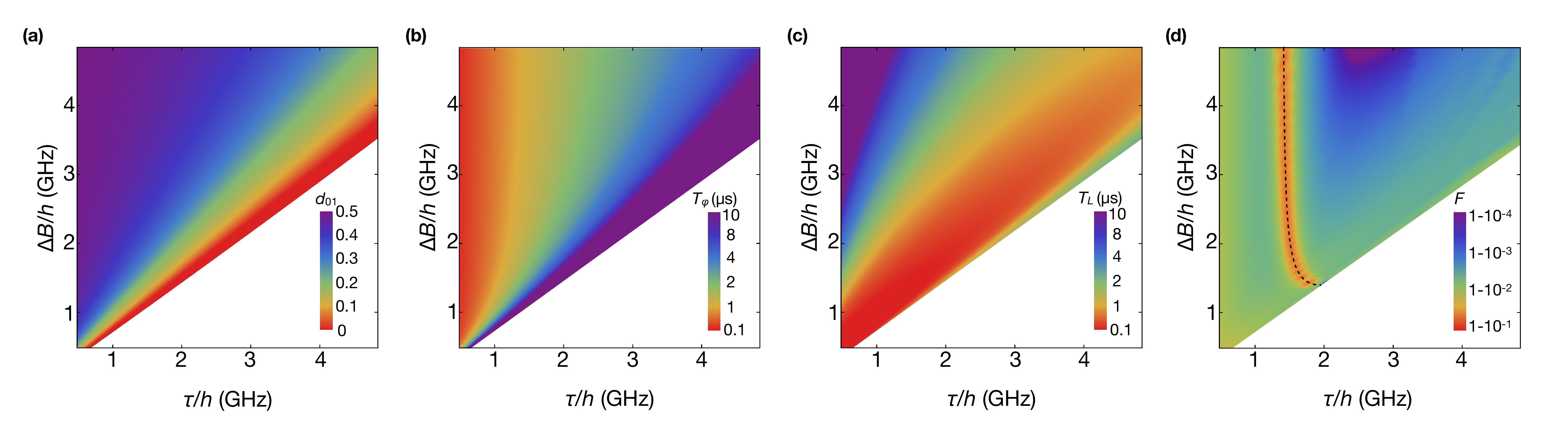}
\caption{Qubit properties, evaluated at a TSS, for a range of $\tau$ and $\Delta B$. 
The large white triangles in the lower-right portions of the plots correspond to regions where $\tau>\tau_\text{crit}$, and no TSS solutions exist.
\textbf{a} Dimensionless dipole moment $d_{01}$, which determines the strength of the transverse coupling. 
When $\tau\ll \Delta B$, $d_{01}\approx 0.5$ because the TSS occurs near the $S(1,1)$-$S(0,2)$ transition where the qubit is charge-like, while at $\tau=\tau_{\rm crit}$, $d_{01}=0$. 
\textbf{b}-\textbf{d} Effects of $1/f$-type charge noise. 
\textbf{b} Dephasing times.
Here, $T_\varphi$ is mainly determined by the width of the sweet spot, which is minimized for TSS near the charging transition (Fig.~2b).
\textbf{c} Leakage excitation times. 
Leakage is maximized when $\tau\approx \Delta B$, because $E_{1L}\approx E_{01}$ (Fig.~2c).
\textbf{d} AC-driven, single-qubit $X_\pi$ gate fidelity.
The dashed line corresponds to the qubit-cavity resonance condition, $\omega_q=\omega_r$; fidelity is suppressed near this line due interference between the AC drive and the cavity mode.
Here we assume a constant photon decay rate of $\kappa/2\pi=0.028$~MHz in the cavity (corresponding to a resonator quality factor of $Q=10^5$)\citep{SamkharadzePRA2016}.
We also assume physically realistic driving and resonator parameters given by $\varepsilon_\text{AC}/h=0.5$~GHz\citep{Brandurnpj2017}, $g_0/2\pi= 50$~MHz\citep{MiNature2018}, $\omega_r/2\pi=2.8$~GHz\citep{SamkharadzePRA2016}, and a cavity temperature of 20~mK for the initial photon state\citep{BenitoPRB2017}. 
}
\label{fig3}
\end{figure*} 

A typical energy level diagram for $H_\text{ST}$ is shown in Fig.~2a for $\tau\gtrsim\Delta B$ (solid lines); here we have assumed large values of $\tau$ and $\Delta B$ to help visualize the key features of the plot. 
In this regime, the qubit energy splitting, $\hbar\omega_q=E_1-E_0$, has only one sweet spot, located at the SOP (Fig.~2b), and the qubit states are largely unperturbed from $\ket{S(1,1)}$ and $\ket{T_0(1,1)}$.
In contrast, when $\tau$ falls below a critical value, $\tau_\text{crit}\approx 1.37 \Delta B$, a dip emerges in the energy dispersion near $\varepsilon=0$, representing a new sweet spot -- the TSS. 
In this case, the energy levels are strongly hybridized and bent (dashed lines in Fig.~2a), yielding eigenstates that resemble $\ket{\uparrow\downarrow}$ and $\ket{\downarrow\uparrow}$.
(Note that mirror-symmetric features are also observed near $\varepsilon = -2U$.
However, since we focus here on the regime near $\varepsilon=0$, the magnitude of $U$ and the presence of $\ket{S(2,0)}$ have almost no effect on the results reported below.
For convenience, we therefore set $U=3$~meV and ignore it for the remainder of this work.)
For $\tau\leq\tau_\text{crit}$, the curvature of $\hbar\omega_q$ is initially positive at the SOP, implying the existence of three sweet spots (ignoring the mirrored sweet spots): the broad valley at the SOP, the much narrower TSS dip near $\varepsilon=0$, and a very small peak in-between (Fig.~2b, lower inset), which we refer to as the alternative transverse sweet spot (ATSS).
As its name indicates, the ATSS also has a transverse coupling, and its position on the $\varepsilon$ axis may occur anywhere between the SOP and the TSS, depending on the value of $\tau$ (Fig.~2b, upper inset).
At a second critical value of $\tau$, $\tau_\text{SSS}\approx\sqrt{3/2}\Delta B\approx 1.22\Delta B$ (the ``super sweet spot"), the ATSS merges with the SOP.
For $\tau<\tau_\text{SSS}$, the curvature of the energy dispersion at the SOP becomes negative, and only two sweet spots remain -- the SOP and the TSS.

The extent to which the TSS, ATSS, and SOP sweet spots are protected from charge noise depends on the flatness of the energy dispersion, which is determined in part by the order of the sweet spot: a sweet spot is classified as $n^\text{th}$-order if $\partial^m \omega_q/\partial \varepsilon^m =0$ for all $m\leq n$.
The SOP is a first-order sweet spot.
However in Supplementary Sec.~S1\citep{SM}, we show that higher derivatives of the energy dispersion can be very small, in terms of the parameter $(\Delta B/U)^m\ll 1$, yielding an approximate ninth-order sweet spot when $\tau=\tau_\text{SSS}$ (and an exact third-order sweet spot), which accounts for the extreme flatness of the energy dispersion.

While single-qubit gates can be performed at the SOP, using the tunnel coupling $\tau$ as a control parameter\citep{ReedSOP,Martins:2016}, the absence of a charge dipole moment makes it more difficult to implement resonator-mediated gates\citep{HarveyPRB2018}.
On the other hand, for the same reason, the SOP makes a useful idling point for qubits coupled to a cavity. 
At the special point, $\tau=\tau_\text{SSS}$, the extreme flatness of the energy dispersion makes the SOP an excellent idling point.
The dipole moment of the ATSS is also small, as discussed below, and the sweet spot is relatively broad, making it an alternative candidate for idling; however the position of the ATSS varies rapidly as function of $\tau$ (Fig.~2b, upper inset), which could present a challenge for tuning.
The TSS forms a narrower sweet spot (Fig.~2b), and its charge dipole is large, which increases its sensitivity to charge noise, but makes it a good candidate for performing gate operations and coupling to a cavity.
In principle, it is possible to adiabatically transition between the TSS and the ATSS, and then the SOP, by simultaneously adjusting the parameters $\tau$ and $\varepsilon$ (Fig.~2b, upper inset), even while $\Delta B$ remains fixed, as is typical in a given experiment. 
We now explore these possibilities in greater detail.

\textbf{Characterizing the TSS}
The position of the TSS in detuning space, $\varepsilon_\text{SS}$, depends on all the parameters of the Hamiltonian, but generally occurs near $\varepsilon=0$ (Fig.~2b, upper inset).
As shown below, the location of the operating point plays a key role in determining the qubit behavior, which has two basic types.
(1) When $\tau\gtrsim\Delta B$ (Fig.~2c), we mainly find that $\varepsilon_\text{SS}<0$; in this case, the energy splitting of the lowest non-logical state $\ket{S(0,2)}$ is approximately resonant with the qubit frequency, resulting in enhanced leakage.
(2) When $\tau<\Delta B$ (Fig.~2d), we have $\varepsilon_\text{SS}>0$; in this case, leakage is suppressed, but the TSS is very narrow, and the qubit is charge-like.
More generally, any qubit property (e.g., decoherence, coupling, or gate fidelity) depends on the specific control parameters.
We now evaluate and compare these properties, first for an isolated qubit, then for a qubit coupled capacitively to a microwave resonator. 

We first consider single-qubit gate operations in isolated qubits.
The gates are performed by applying an AC drive to the detuning parameter. In the presence of charge noise $\delta \varepsilon(t)$, the time-dependent detuning is given by $\Delta \varepsilon(t) = \varepsilon_\text{AC} \cos(\omega t) + \delta \varepsilon(t)$.
From Eq.~(1), the resulting interaction is given by
\begin{equation}
H_{\rm int}=\Delta\varepsilon(-|S(0,2)\rangle\langle S(0,2)|+|S(2,0)\rangle\langle S(2,0)|).
\end{equation}
Since the states $\ket{S(0,2)}$ or $\ket{S(2,0)}$ generate the charge dipole in this system, $H_{\rm int}$ is proportional to the dimensionless dipole operator, $\hat d=\partial H_\text{ST}/\partial \varepsilon$.
In general, $H_\text{int}$ can have longitudinal and transverse components; however at a sweet spot, the longitudinal component vanishes, by definition.

We begin by solving the total Hamiltonian, defined as $H=H_\text{ST}+H_\text{int}$.
First, we ignore the state $\ket{S(2,0)}$ in Eq.~(1), since it is very high in energy.
We then evaluate $H$ in the $\{\ket{0},\ket{1},\ket{L}\}$ eigenbasis, which diagonalizes $H_\text{ST}$, obtaining
\begin{equation}
H_{q}= \sum_n E_n\sigma_{nn}+\Delta\varepsilon\sum_{n,m} d_{nm}\sigma_{nm} .
\label{eq:Hqr}
\end{equation}
Here $E_n$ are the eigenvalues of $H_\text{ST}$, $\sigma_{nm}=\ket{n}\bra{m}$, where $n,m \in \{0,1,L\}$, and $d_{nm}=\bra{n}\hat d\ket{m}$, where $\Delta\varepsilon\, d_{01}$ is the transverse dipole coupling induced by $H_\text{int}$.
In Fig.~3a, we plot numerical solutions for $d_{01}$, evaluated at the TSS, as a function of $\tau$ and $\Delta B$.
The large white triangle in the lower-right portion of the plot corresponds to $\tau>\tau_\text{crit}(\Delta B)$, where no TSS solutions exists.
The general features of the plot can be understood as follows.
When $\tau\ll\Delta B$, the TSS occurs near the charging transition, which causes $\ket{S(0,2)}$ and $\ket{S(1,1)}$ to strongly hybridize, and yields an effective charge qubit, for which $d_{01}\approx 0.5$ at the sweet spot.
When $\tau\rightarrow \tau_\text{crit}$, the TSS moves away from the charging transition, resulting in a suppressed dipole, $d_{01}\approx 10^{-3}$-$10^{-4}$, which vanishes completely at the SOP.
To a good approximation, $d_{01}$ depends only on the ratio $\tau/\tau_\text{crit}\propto \tau/\Delta B$ over the entire plot range of Fig.~3a, yielding a radial plot.
The large-$d_{01}$ (small-$\tau/\Delta B$) operating regime is preferential for boosting gate speeds; however we now show that the dephasing rate $1/T_\varphi$ also grows in this regime.

We define $T_{\varphi}$ as the decay time of the $\rho_{01}$ component of the qubit density matrix, which we estimate by simulating its free-induction decay.
As described in Methods, we introduce $\delta\varepsilon(t)$ fluctuations into the simulations, sampling from a gaussian distribution with $1/f$ spectral correlations.
We then average over a large number of charge-noise realizations to obtain the results shown in Fig.~3b.
Since the TSS is a sweet spot, it is protected from small $\delta\varepsilon$ fluctuations, to lowest order.
The main contribution to dephasing therefore occurs at order $\delta\varepsilon^2$, and  its behavior correlates with the width of the sweet spot.
In the $\tau\lesssim\Delta B$ regime (Fig.~2d), the TSS is well separated from other features in the energy dispersion; its shape therefore does not depend on $\Delta B$, which only determines the splitting between states $\ket{\uparrow\downarrow}$ and $\ket{\downarrow\uparrow}$.
Hence, the qubit is charge-like, and the width of the sweet spot is determined by $\tau$ rather than $\Delta B$.
This is consistent with Fig.~2b where the sweet spot is quite narrow for small $\tau$.
It is also consistent with Fig.~3b where $T_\varphi$ approaches 100~ns in the limit $\tau\ll\Delta B$, and becomes independent of $\Delta B$.
On the other hand, for $\tau\rightarrow \tau_\text{crit}$, we observe a wider sweet spot in Fig.~2b.
In this regime, the presence of the leakage state actually helps to flatten the energy dispersion, yielding $T_\varphi$ approaching 10~$\mu$s.

Using the same simulations, we also compute $T_L$, defined here as the decay time of $\rho_{00}(t)+\rho_{11}(t)$, due to leakage. The results, which are plotted in Fig.~3c, exhibit a similar range of timescales as Fig.~3b; however, the trends are very different.
This is easy to understand because leakage is caused by the hybridization of logical and non-logical state(s), which occurs near the resonance condition $E_L-E_1=E_1-E_0$, causing a dip in $T_L$ when $\tau\approx \Delta B$.

Finally, we note that phonon-mediated decay processes have not been considered in the current analysis, although they also contribute to $T_1$-type relaxation.
For GaAs-based devices, such processes are expected to reduce $T_1$ to a few nanoseconds for the large magnetic field gradients considered here, due to the presence of piezoelectric phonons\citep{Kornich2018phononassisted}.
In the current proposal, we have therefore focused on Si-based devices, where piezoelectric phonons are absent, and the phonon-mediated $T_1$ is generally much longer than any time scale relevant to our analysis\cite{PrancePRL2012, Kornich2018phononassisted}.

To summarize the results of this section, the behaviors of $T_\varphi$ and $T_L$ exhibit opposite trends as a function of $\tau$ when $\varepsilon$ is tuned to a TSS; the best working points must therefore be determined via optimization. 
We address this problem below, by computing the fidelities of one- and two-qubit gates.

\textbf{Single-Qubit Gate Fidelity.}
In the previous section, we studied free induction.
Here we consider resonantly driven, single-qubit $X_\pi$ gate operations performed at a TSS.
We consider single-qubit interactions mediated by AC-driven gates, which are generally expected to be faster than single-qubit gates mediated by a resonator.
However, the two-qubit gates in the following section are mediated by a resonator, with a capacitive coupling that cannot be turned off, as indicated in Fig.~1; we therefore include this interaction in the present analysis. 
In our simulations, we further assume that the cavity resonant frequency $\omega_r$ cannot be tuned. 
However, we note that the cavity-qubit detuning $\Delta_0 = \hbar  (\omega_r- \omega_q)$ can be varied, because the qubit frequency depends on the parameters $\Delta B$ and $\tau$.

We model the qubit-resonator system with the Hamiltonian 
\begin{multline}
H_{qr}=\hbar\omega_ra^\dagger a + \sum_n E_n\sigma_{nn} \\ 
+\sum_{n,m} \left[ \Delta\varepsilon\, d_{nm}\sigma_{nm}+\hbar g_0d_{nm}(a+a^\dagger)\sigma_{nm} \right],
\label{eq:Hqr}
\end{multline}
where $a^\dagger(a)$ is the photon creation (annihilation) operator, $g_0=eV_0$ is the bare capacitive coupling between the qubit and resonator, $V_0=\sqrt{\hbar Z_r}\,\omega_r$ is the amplitude of the resonator voltage anti-node, and $Z_r$ is the resonator impedance\citep{HarveyPRB2018}. 
The effective qubit-cavity coupling, $g=g_0 d_{01}$, is proportional to the transverse dipole moment, which is maximized near the charging transition.
As noted above, the coupling can be turned off ($d_{01}=0$) at the SOP, while $d_{01}\approx 0.5$ for large $\Delta B$.

We perform simulations of Eq.~(\ref{eq:Hqr}) for a range of $\Delta B$ and $\tau$.
For each pair of values, $(\Delta B,\tau)$, we tune the intra-qubit detuning parameter to a TSS [$\varepsilon=\varepsilon_\text{SS}(\Delta B,\tau)$] to improve the gate fidelity, and apply an AC drive at the qubit resonant frequency, $\Delta\varepsilon(t)=\varepsilon_\text{AC}\cos(\omega_qt)$. 
Since we do not limit the simulations to the weak-driving regime, the $X_\pi$ gate times must be determined numerically; we do this by evolving over many Rabi oscillations, to more accurately locate the initial peak.
The simulations are computationally expensive, compared to Figs.~3a and 3b, since they include photon basis states. Therefore, we do not explicitly include either charge noise or photon decay at the Hamiltonian level. 
Instead, we solve a master equation based on Eq.~(\ref{eq:Hqr}), in which dephasing effects are included phenomenologically through the dephasing rate $1/T_\varphi$ and leakage effects are included through the decay rate $1/T_L$, which were both obtained as functions of $\Delta B$ and $\tau$ in the previous section. 
Resonator photon decay is included through a constant decay rate, $\kappa$.
We then compute the gate fidelity, obtaining the results shown in Fig.~3d.
See Methods for details of these calculations.

We observe the following behavior.
First, gate fidelities are generally found to be high, except very near the resonance condition $\Delta_0=0$ (dashed line), where excited photons in the resonator form leakage levels that naturally suppress the single-qubit gate fidelity.
For larger values of $\kappa$, the fidelity is further suppressed near the resonance condition.
On the other hand, $\Delta_0$ increases quickly as we move away from this line, suppressing this effect.
Even further away from the resonance condition, the gate fidelity is slightly suppressed for small $\tau$, due to strong dephasing (Fig.~3b), or near the line $\tau=\tau_\text{crit}$, due to enhanced leakage (Fig.~3c) and smaller charge dipoles (Fig.3a).
The best fidelities are therefore obtained midway between the resonance condition and $\tau=\tau_\text{crit}$, at larger values of $\Delta B$.
For the physically realistic simulation parameters used in Fig.~3d, the fidelities can be quite high, approaching 99.99\%.
Finally, we note that closer inspection of the resonance condition in Fig.~3d reveals weak oscillations.
As discussed in Supplementary Sec.~S2, these can be understood as a combination of leakage and strong-driving effects.

\textbf{Two-Qubit Gate Fidelity.}
We consider two-qubit gates mediated by a cavity, with a set-up similar to Fig.~1, and with both qubits positioned at voltage anti-nodes.
Simulations are performed analogously to the previous section, but with a two-qubit Hamiltonian given by
\begin{multline}
H_{qqr}= \hbar\omega_r a^\dagger a +\sum_{i=a,b}\sum_n E_{n,i}\sigma_{nn,i} \\ 
+\sum_{i=a,b}\sum_{n,m}\hbar g_{nm,i}(a+a^\dagger)\sigma_{nm,i},
\label{eq:Hqqr}
\end{multline}
where $g_{nm,i}=g_0d_{nm,i}$ and the subscript $i$ refers to qubits $a$ or $b$.
In the dispersive limit, the native gate for $H_{qqr}$ is iSWAP, with gate times determined analogously as for single-qubit gates.
The two-qubit gate can be switched off by tuning either of the qubits to its SOP.
To simplify the following analysis, we set $\Delta_{0a}=\Delta_{0b}\equiv\Delta_0$, $g_{nm,a}=g_{nm,b}\equiv g_{nm}$, $\Delta B_a=\Delta B_b\equiv\Delta B$, $\tau_a=\tau_b\equiv \tau$, and $\varepsilon_a=\varepsilon_b\equiv\varepsilon$, to reduce the number of independent control parameters.
The gate fidelities are computed by solving the master equation associated with Eq.~(\ref{eq:Hqqr}), including the decoherence rates $1/T_L$, $1/T_\varphi$, and $\kappa$, as before, and comparing the result to an ideal iSWAP gate.
Our results are shown in Fig.~4a, using the same simulation parameters as Fig.~3d.

\begin{figure}[t]
\includegraphics[width=3.4 in]{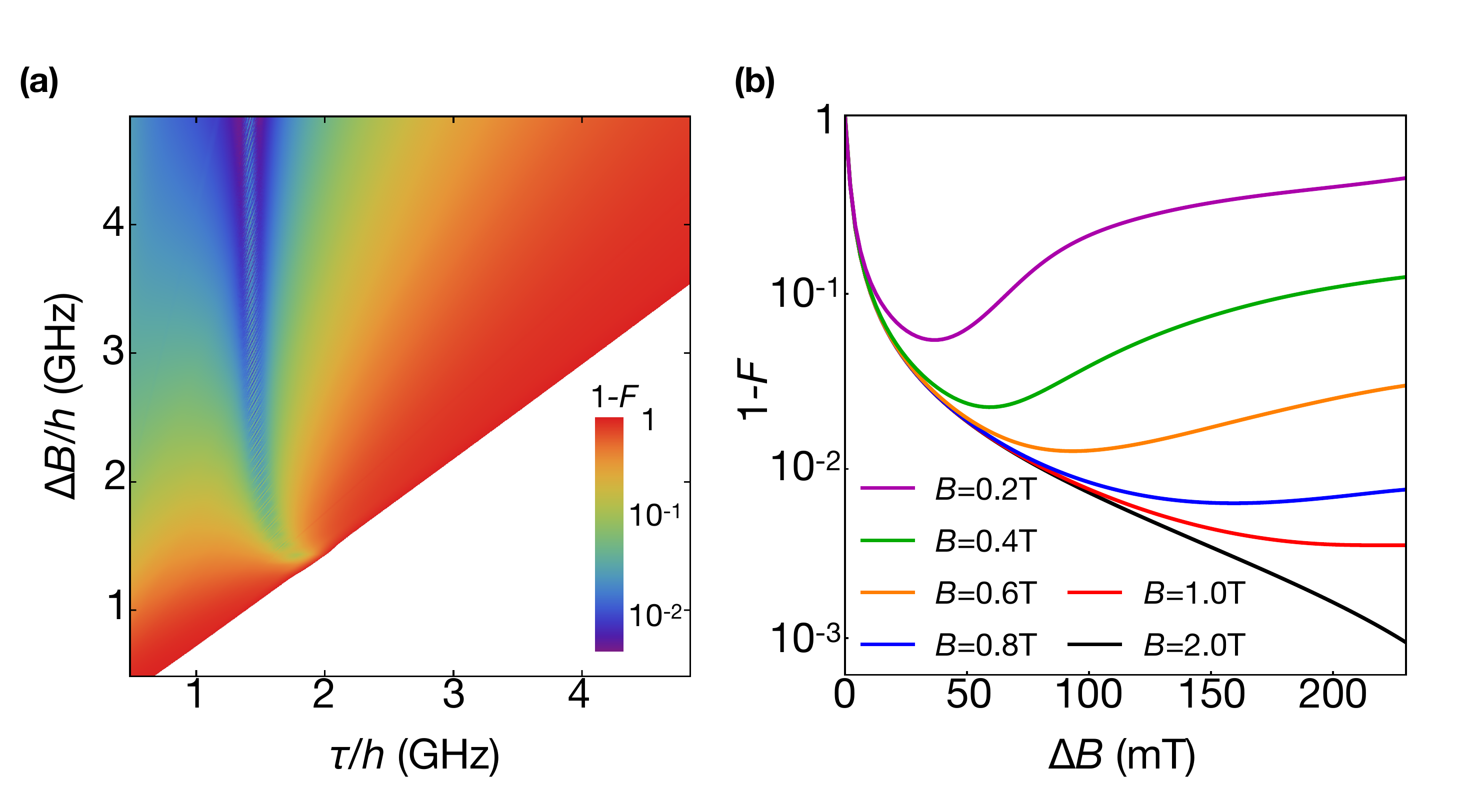}
\caption{
Resonator-mediated two-qubit iSWAP gates, performed at a TSS.
\textbf{a} Gate infidelity, assuming the same device parameters as Fig.~3. 
Optimal fidelities are obtained very near, but on either side of the qubit-cavity resonance condition, near where $d_{01}$ is maximized. 
Here we assume no coupling to spin-polarized leakage states.
\textbf{b} Gate infidelity, including coupling to spin-polarized leakage states.
For each value of $\Delta B$, we plot the maximum fidelity with respect to $\tau$ . 
The leakage coupling is suppressed, and the fidelity is maximized, by applying large global $B$ fields.
}
\label{fig4}
\end{figure} 

Although similar physics determines the fidelities of one and two-qubit gates, the trends observed in Figs.~3d and 4a are very different.
In particular, the fidelity dip along the resonance line in Fig.~3d becomes a double peak in Fig.~4a. 
This is because the single-qubit gates are driven, with the resonator acting only as a leakage channel.
For two-qubit gates, the cavity mediates the interaction, and the fidelity is generally enhanced near the resonance condition, $\Delta_0=0$, where the effective qubit-qubit coupling is maximized\citep{SrinivasaPRB2016}.
(The same is true for single-qubit gates mediated by a resonator, although we do not explore that possibility here.)
Very near the resonance, however, spontaneous excitation of the qubits by the cavity (the Purcell effect) suppresses the gate fidelity (i.e., increases the infidelity), causing maxima to form on either side of this line.
The same process also reduces the individual qubit lifetimes. 
In cases where the Purcell effect dominates the fidelity, we note that an alternative approach would be to replace the cavity with a direct capacitive coupling\citep{NicholNPJQI2017}, although we do not explore that possibility here.

Far from the resonance condition, two-qubit gate fidelities are typically low, because off-resonant gates tend to be slow, and therefore susceptible to charge noise.
(This is not a problem for single-qubit gates, which can be strongly driven.)
However, fidelities are found to increase for larger $\Delta B$, due to stronger qubit-cavity couplings and reduced leakage (Fig.~3).
To exploit this trend, we note that nanomagnets in recent double-dot experiments have already achieved $\Delta B$ values as large as 80~mT\citep{YonedaAPE2015} ($=2.2$~GHz), corresponding to a maximum fidelity of $98.5$\% in Fig.~4a. 
Finally, we note that small fidelity oscillations are observed near the resonance condition, which are reminiscent of those in Fig.~3d, and can also be attributed to leakage and strong driving (Supplementary Sec.~S3).

\textbf{Leakage Induced by the Polarized Triplets.}
Up to this point, we have not considered the polarized spin triplet states, $\ket{\uparrow\uparrow}$ and $\ket{\downarrow\downarrow}$, which present new leakage channels.
In this case, hybridization with the qubit states is caused by a transverse magnetic field gradient.
It is reduced, however, when the levels are split off by a large global field; further details of these calculations are presented in Supplementary Sec.~S4.
To estimate the effect of such leakage on two-qubit gate fidelities, we first extend Eq.~(\ref{eq:Hqqr}) to include a global $B$ field and a transverse field gradient $\Delta B_\perp$.
Since $\Delta B_\perp$ and $\Delta B$ are expected to be similar in size\citep{YonedaAPE2015}, we simply set $\Delta B_\perp=\Delta B$.
We then compute the iSWAP gate fidelity for a fixed $\Delta B$, and determine its maximum as a function of $\tau$.
Repeating this procedure as a function of $\Delta B$, for several values of $B$, yields the results shown in Fig.~4b.
As expected, we find that fidelities improve uniformly as a function of $B$.
However, the dependence on $\Delta B=\Delta B_\perp$ is non-monotonic: the fidelity initially increases (infidelity decreases) by the same mechanism as Fig.~4a; for larger $\Delta B_\perp$, this behavior saturates, and leakage eventually dominates the fidelity.
For the range of $\Delta B$ plotted here, we find that $B\geq 2$~T is sufficient for avoiding most leakage.
More generally, $B \geq 0.8$~T yields fidelities $>99\%$, when $\Delta B > 100$~mT.

\vspace{.1in} \noindent 
\textbf{DISCUSSION} 
\vspace{.05in} \\
We have shown that qubit coherence and one- and two-qubit gate fidelities are strongly affected by the operating points in a control space spanned by the parameters $B$, $\Delta B$, $\Delta B_\perp$, $\tau$, $\varepsilon$, $\varepsilon_\text{AC}$, $g_0$, and $\omega_r$, as well as the noise characteristics of the qubits and the resonator.
The transverse sweet spots (TSS) studied in this work make good working points, because they provide protection against environmental noise while offering a strong coupling to external driving fields or a microwave resonator.

To achieve high-fidelity gates at a TSS, it is important to provide a large gradient-induced Zeeman splitting, $\Delta B$, and a nearly resonant coupling between the qubit and cavity.
Since $\hbar\omega_q\leq2\Delta B$ for a TSS, we therefore require that $2\Delta B\gtrsim\hbar\omega_r$, while noting that neither $\Delta B$ nor $\omega_r$ is easy to change after a device is fabricated.
Fortunately, recent work shows that it is possible to form high-kinetic-inductance resonators with low resonant frequencies, $\omega_r/2\pi\approx 2.8$~GHz\citep{SamkharadzePRA2016},  while maintaining a high cavity $Q>10^5$ in the presence of a large in-plane field $B=6$~T.
Moreover, as noted above, large gradients, $\Delta B/h\approx 2.2$~GHz ($= 80$~mT), have already been achieved in the lab\citep{YonedaAPE2015}, indicating that the requirements for a TSS have already been met.

Adopting the values for $\Delta B$ and $\omega_r$ from the previous paragraph, and choosing $\Delta B_\perp=\Delta B$, $B=2$~T, a resonator coupling of $g_0/2\pi = 0.05$~GHz, and a realistic driving field of $\varepsilon_\text{AC}/h=0.5$~GHz, we obtain the following results at a TSS.
Single-qubit gates are found to be fairly fast, with a gate time of $t_\text{1Q}\approx 7$~ns, yielding a gate fidelity of 99.6\% for an optimal tunnel coupling of $\tau/h=$2.1~GHz.
Two-qubit gates are slightly slower, with $t_\text{2Q}\approx 50$~ns, yielding a gate fidelity of 98.5\% for the optimal tunnel coupling $\tau/h=$1.5~GHz. 
Our simulations also show that when $2\Delta B\simeq \omega_r$, leakage tends to dominate the infidelity, while for smaller values of $\Delta B$, dephasing is the dominant problem.
We find that, as a rule of thumb, $1.5\Delta B\simeq \omega_r$ provides a good balance for obtaining higher fidelities, which explains our choice of $\Delta B$ and $\omega_r$ values in the simulations.
However, better fidelities can be achieved by increasing both of these parameters simultaneously.
Theoretical calculations suggest that larger $\Delta B\geq 150$~mT values should be possible for near-term experiments\cite{LamarreAIP2013,ChesiPRB2014}.
Repeating our simulations with this $\Delta B$ and $\omega_r/2\pi =5.2$~GHz gives optimal fidelities of 99.95\% and 99.6\% for one and two-qubit gates, respectively.

The results described above exploit optimized TSS working points for singlet-triplet qubits, but reveal that these points differ for single and two-qubit gate operations.
In addition, resonator-mediated gates require an idling point, where the effective coupling to the resonator is turned off.
We have identified two good candidates for idling points: the ATSS, where $d_{01}$ is very small, or the SOP where $d_{01}=0$, particularly when $\tau=\tau_\text{SSS}$.
Fortunately, for a fixed value of $\Delta B$, we can navigate between gating and idling points while maintaining a TSS, by simultaneously tuning $\tau$ and $\varepsilon$ such that $\varepsilon=\varepsilon_\text{SS}(\tau)$.

Longitudinal and transverse couplings can be viewed as distinct, physical resources, with unique advantages and disadvantages for quantum computing.
It is therefore important to compare their attributes\citep{LambertPRB2018}; the singlet-triplet qubit provides a testbed for doing so in a single experimental setting.
In this work, we have focused on the TSS, which has a purely transverse coupling and can be formed over a continuous range of parameters.
In fact, the TSS and ATSS are the only tunings with purely transverse couplings for singlet-triplet qubits.
The SOP is the only tuning with a purely longitudinal, ``curvature"-type coupling (see below), which can be formed over a continuous range of $\tau$ when $\varepsilon=-U$.
All other operating points have both transverse and longitudinal components.
Such mixing reduces the response to AC driving for single-qubit gates, and yields complicated behavior for two-qubit gates, which may be undesirable from a control perspective.
These mixed operating points also do not correspond to sweet spots, and should therefore experience faster decoherence.
The TSS coupling is particularly strong because the qubit's charge character is maximized.
In contrast, at the SOP, the charge dipole vanishes, resulting in a weaker, second-order ``curvature" coupling\citep{Ruskov2017}, which is consistent with slower gates that are well protected by a high-order sweet spot.
Alternatively, gate speeds at the SOP may be enhanced by employing AC driving techniques\citep{HarveyPRB2018}.
A thorough experimental comparison of longitudinal vs.\ transverse couplings should consider all of these possibilities.

\vspace{.1in} \noindent 
\textbf{METHODS} 
\vspace{.05in} \\
In this work, we perform two types of numerical simulations: (i) free induction of single qubits, and (ii) one and two-qubit gate operations.
The simulations employ different theoretical methods, and are repeated for cases with and without charge noise.
All numerical calculations use the QuTiP software package\citep{QuTiP}.

Free-induction simulations are performed after adding time-dependent charge noise to the detuning parameter in Eq.~(3), with $\Delta\varepsilon=\delta\varepsilon(t)$.
Noise sequences are generated following the method described in refs. \onlinecite{Kawakami11738,Yang:2019}:
we first generate random white noise $\delta\varepsilon(t)$ over a discrete time sequence.
This sequence is then Fourier transformed and scaled in frequency space by the noise power spectrum $\sqrt{S(\omega)}$, where
\begin{equation}
\label{eq:noise}
{S}(\omega) = c_{\varepsilon}^2 \left\{
  \begin{array}{cl}
    \frac{2\pi}{|\omega|} & \text{for}\,\, \omega_l \leq |\omega | \leq \omega_h\\
    0 & \text{otherwise} 
  \end{array}
\right. ,
\end{equation}
and $\omega_l/2\pi=$100~kHz and $\omega_h/2\pi=$20~GHz are lower and upper frequency cutoffs.
We choose a noise strength of $c_\varepsilon=0.56$~$\mu$eV, corresponding to a standard deviation of $\sigma_\varepsilon=2$~$\mu$eV for noise integrated over the entire frequency spectrum, as consistent with several recent experiments{\citep{jock2018silicon, Brandurnpj2017, ShiPRB2013}.
The resulting frequency sequence is Fourier transformed back to the time domain, yielding the desired noise sequence.
For each point in the $\Delta B$-$\tau$ plots shown in Figs.~3b and 3c, we average the density matrix $\rho(t)$ over 10,000 different noise realizations, with initial states $\rho(0)=\ket{i}\bra{i}$, where $\ket{i}=(\ket{0}+\ket{1})/\sqrt{2}$.
We use the same simulations to obtain the density matrix $\rho_\text{leak}(t)$, using  $\rho_\text{leak}(0)=(\ket{0}\bra{0}+\ket{1}\bra{1})/2$, to obtain purely leakage errors. 
$T_\varphi$ and $T_L$ are obtained by fitting the averaged results to\citep{HuFoundations}
\begin{equation}
|\rho_{01}(t)|=|\rho_{01}(0)|\exp\left[-(t/T_\varphi)^\beta\right]
\end{equation}
and
\begin{equation}
\rho_{00}(t)+\rho_{11}(t)=\frac{1}{3}\exp(-t/T_L)+\frac{2}{3},
\end{equation}
where $\rho=\mathbf{1}/3$ represents the fully mixed state in the $\{\ket{0},\ket{1},\ket{L}\}$ basis. 
$\beta$ is also left as a fitting parameter, to account for the fact that non-dephasing, leakage processes can dominate the decoherence in some cases.
Here, we only assume coupling to the dominant leakage state associated with $\ket{S(0,2)}$.

For one and two-qubit gates, we incorporate the free-induction results into our simulations of the qubit-cavity master equation, defined as\citep{SrinivasaPRB2016}
\begin{multline}
\dot{\rho}=-\frac{i}{\hbar}[H,\rho]+\frac{\kappa}{2}(2a\rho a^\dagger-a^\dagger a\rho-\rho a^\dagger a)\nonumber \\ 
+\sum_j \Big[\frac{1}{2T_\varphi}(\sigma_{z,j}\rho\sigma_{z,j}-\rho)+ \nonumber \\ 
\frac{1}{2T_L}(2\sigma_{L,	j}\rho\sigma_{L,j}^{\dagger}-\sigma_{L,j}^{\dagger}\sigma_{L,j}\rho-\rho\sigma_{L,j}^{\dagger}\sigma_{L,j})\Big] ,
\end{multline}
where $H$ represents the appropriate one-qubit ($j=a$) or two-qubit ($j=a,b$) Hamiltonian, given in Eqs.~(4) or (5) of the main text, $\kappa$ is the cavity decay rate, $T_\varphi$ and $T_L$ are computed as functions of $\Delta B$ and $\tau$, as described above, $\sigma_{z,i}=\sigma_{11,i}-\sigma_{00,i}$ is the dephasing operator for qubit $i$, and $\sigma_{L,i}=\sum_{n=0,1} \sigma_{nL,i}$ is the operator associated with leakage between the logical subspace of qubit $i$ and its leakage state $L$.
Initial states for the calculations are given by $\rho(0)=\ket{i}\bra{i}$, where $\ket{i}=\ket{0}_c(\ket{0}+\ket{1})/\sqrt{2}$ for single-qubit gates (here, $\ket{0}_c$ represents the zero-photon state of the cavity), or $\ket{i}=\ket{0}_c\ket{e}_a\ket{g}_b$ for two-qubit gates, where $\ket{e}_a$ and $\ket{g}_b$ represent the ground and excited qubit eigenstates for qubits $a$ and $b$, respectively (e.g., see ref. \onlinecite{SrinivasaPRB2016}).
The corresponding gate fidelities are then computed from
\begin{equation}
F={\rm Tr}[\rho(t_g)\rho_{\rm ideal}(t_g)], \label{eq:fid}
\end{equation}
where $t_g$ is the appropriate gate time, and the ideal density matrix is computed in the absence of noise or leakage-state couplings.

\vspace{.1in} \noindent 
\textbf{DATA AVAILABILITY} 
\vspace{.05in} \\
Data sharing is not applicable to this article as no data sets were generated or analyzed during the current study. The simulation results reported may be obtained by following the computational scheme described in Methods. All numerical calculations use the QuTiP software package\citep{QuTiP}. The Mathematica code for the characterization of the SOP is provided as Supplemental Material.

\vspace{.1in} \noindent 
\textbf{ACKNOWLEDGMENTS} 
\vspace{.05in} \\
We are grateful to Mark Gyure, Joseph Kerckhoff, Thaddeus Ladd, and Emily Pritchett for illuminating discussions.
This work was supported in part by the Army Research Office (W911NF-17-1-0274) and the Vannevar Bush Faculty Fellowship program sponsored by the Basic Research Office of the Assistant Secretary of Defense for Research and Engineering and funded by the Office of Naval Research through Grant No.\ N00014-15-1-0029. The views and conclusions contained in this document are those of the authors and should not be interpreted as representing the official policies, either expressed or implied, of the U.S. Government. The U.S. Government is authorized to reproduce and distribute reprints for Government purposes notwithstanding any copyright notation herein.

\vspace{.1in} \noindent 
\textbf{COMPETING INTERESTS} 
\vspace{.05in} \\
The authors declare that there are no competing interests.

\vspace{.1in} \noindent 
\textbf{AUTHOR CONTRIBUTIONS} 
\vspace{.05in} \\
J.C.A.-U. performed numerical simulations and analytical calculations.   J.C.A.-U.,  M.E., S.N.C.,  and  M.F.  analyzed  the  results  and  prepared  the  manuscript.   Work  was  carried out under the supervision of S.N.C. and M.F.

\beginsupplement
\pagebreak
\begin{widetext}

\section{Characterizing the SOP sweet spot}
In the main text, we discuss the order of the sweet spot at the symmetric operating point (SOP).
This requires evaluating the qubit frequency $\omega_q$ and its derivatives $d^n\omega_q/d\varepsilon^n$ at the sweet spot.
To simplify the derivation, $\omega_q$ can be expanded in terms of the small parameters $\varepsilon/U$, $\tau/U$, and $\Delta B/U$.
Here we explain how the calculations are performed using Mathematica symbolic manipulation software\cite{Mathematica}. 
Since the resulting analytical expressions are complicated, we outline the steps of the calculations here, and provide details in the accompanying Supplemental Mathematica Notebook.

First, we divide the singlet-triplet Hamiltonian $H_\text{ST}$ [Eq.~(1) of the main text] by $U$, and express all quantities in terms of the dimensionless parameters $\varepsilon/U$, $\tau/U$, and $\Delta B/U$. 
$H_\text{ST}$ is then diagonalized.
The two lowest energy eigenvalues $E_0$ and $E_1$ are identified by evaluating the symbolic expressions at typical values of $\varepsilon$, $\tau$, $\Delta B$, and $U$.
This allows us to construct a formal expression for $\hbar\omega_q=E_1-E_0$.

Second, we expand $f_q=\omega_q/2\pi$ in terms of the small parameters $\varepsilon/U$, $\tau/U$, and $\Delta B/U$.
The resulting series takes the form
\begin{equation}
h f_q(\varepsilon,\tau,\Delta B,U)=U\sum_{i=0} f_i(\tau/U,\Delta B/U)(\varepsilon/U)^i, \label{eq:fQexpand}
\end{equation}
and is expected to converge rapidly for realistic parameter values.

Third, we solve for several $f_i$ by taking appropriate derivatives of $\omega_q$.
Due to the symmetry of the SOP, we note that the odd-order derivatives, $d^{2n-1}\omega_q/d\varepsilon^{2n-1}|_\text{SOP}$, all vanish for $n\geq 1$, yielding $f_{2n-1}=0$.
We also note that the term $f_2(\tau/U,\Delta B/U)$ vanishes when the dispersion curvature changes sign at the super sweet spot $\tau=\tau_\text{SSS}$. 
Solving $f_2(\tau_\text{SSS}/U,\Delta B/U)=0$ then gives $\tau_\text{SSS}/U=\sqrt{3/2}\, \Delta B/U+\mathcal{O}\left[(\Delta B/U)^5\right]$, as reported in the main text.

Finally, after setting $\tau_\text{SSS}=\sqrt{3/2}\, \Delta B$, we can evaluate several terms in Eq.~(\ref{eq:fQexpand}), obtaining
\begin{gather}
f_4(\tau_\text{SSS}/U,\Delta B/U)=0+\mathcal{O}\left[(\Delta B/U)^3\right] , \nonumber \\
f_6(\tau_\text{SSS}/U,\Delta B/U)=0+\mathcal{O}\left[(\Delta B/U)^3\right]\nonumber , \\
f_8(\tau_\text{SSS}/U,\Delta B/U)=0+\mathcal{O}\left[(\Delta B/U)^3\right]\nonumber , \\
f_{10}(\tau_\text{SSS}/U,\Delta B/U)=0+\mathcal{O}\left[(\Delta B/U)^3\right]\nonumber .
\end{gather}
We note, however, that even though $(\Delta B/U)^3\ll 1$, the tenth order term $f_{10}$
has a large prefactor, such that $f_{10}(\tau_\text{SSS}/U,\Delta B/U)>0.1$ for typical parameter values.
Hence, we refer to the SOP with $\tau=\tau_\text{SSS}$ as an approximate ninth-order sweet spot, which explains the extreme flatness of the curvature.

\begin{figure}[t]
\includegraphics[width=5 in]{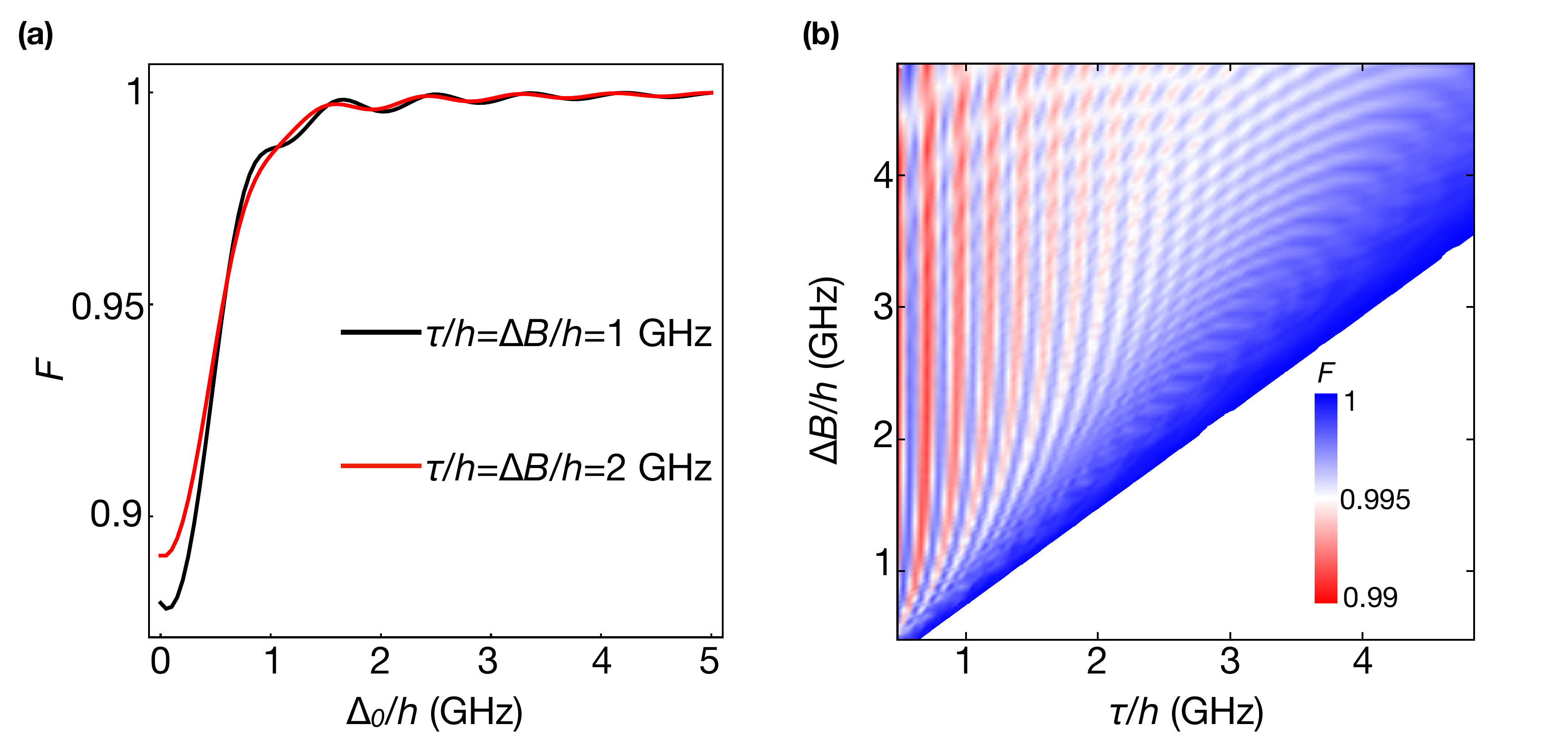}
\caption{AC-driven single-qubit $X_\pi$ gate fidelities in the absence of charge noise. 
\textbf{a} Gate fidelity $F$ as a function of qubit-resonator detuning $\Delta_0$ for two different values of $\tau$ and $\Delta B$ (and hence, $\omega_q$). 
Far off resonance, no photons are excited in the cavity (i.e., no leakage), and the gate fidelity approaches unity. 
Near the resonance condition $\Delta_0=0$, photonic leakage is enhanced.
(Note that the fidelity axis does not extend all the way to $F=0$ here, for visual clarity.)
\textbf{b} Gate fidelity $F$ as a function of tunnel coupling $\tau$ and magnetic field gradient $\Delta B$ at the TSS, for a fixed value of $\Delta_0/h=1.5$ GHz. 
As explained in the main text, the oscillations here can only be attributed to strong-driving effects.}
\label{figS}
\end{figure} 

\section{Single-qubit fidelity oscillations}
In Fig.~3d of the main text we observe weak oscillations of the single-qubit gate fidelity very near the resonance condition, which is indicated as a dashed line.
Here we show that this behavior occurs due to a combination of leakage and strong driving effects.

The full Hamiltonian of the qubit-cavity system is given in Eq.~(4) of the main text as
\begin{equation}
H_{qr}=\hbar\omega_ra^\dagger a + \sum_n E_n\sigma_{nn} 
+ \sum_{n,m} \left[ \Delta\varepsilon d_{nm}\sigma_{nm}+\hbar g_0d_{nm}(a+a^\dagger)\sigma_{nm}
\right] ,
\label{eq:HqrSM}
\end{equation}
where all parameters are described in the main text.
In the following discussion, we can ignore the leakage state $L$ in this expression, since it plays no important role. 
Evaluated at the TSS, Eq.~(\ref{eq:HqrSM}) then reduces to
\begin{equation}
H_{qr}^{(0,1)}=\frac{\hbar\omega_q}{2}\sigma_z+\hbar\omega_ra^\dagger a+\varepsilon_\text{AC} d_{01}\cos(\omega_qt)\sigma_x+\hbar g(a+a^\dagger)\sigma_x,
\end{equation}
where the resonant drive is given by $\Delta\varepsilon=\varepsilon_\text{AC}\cos (\omega_q t)$, and $g=g_0d_{01}$.
We can gain insight into the fidelity oscillations of Fig.~3d by moving to the rotating frame defined by
 \begin{equation}
 H_0=\frac{\hbar\omega_q}{2}\sigma_z+\hbar\omega_ra^\dagger a.
 \end{equation}
The resulting interaction Hamiltonian is given by
\begin{equation}
H_\text{int}=\frac{\varepsilon_\text{AC}}{2}d_{01}\sigma_x
+\hbar g(a^\dagger\sigma_-e^{i\Delta_0 t/\hbar}+a\sigma_+e^{-i\Delta_0 t/\hbar})
+\hbar g(a^\dagger\sigma_+e^{i(\omega_q+\omega_r) t/\hbar}+a\sigma_-e^{-i(\omega_q+\omega_r) t/\hbar}).
\label{eq:Hint1}
\end{equation}
Here, the first term is responsible for Rabi oscillations.
The second term causes unwanted excitations of photons in the cavity, corresponding to leakage, and its effect is maximized at the resonance condition, $\Delta_0=0$.
The third term contains the counter-rotating terms; however both the second and third terms correspond to strong-driving effects.

In Fig.~S1a, we plot the fidelity of an $X_\pi$ gate as a function of the qubit-cavity detuning $\Delta_0$.
(Note that since $\tau$ and $\Delta B$ are fixed for each curve, this corresponds to varying the parameter $\omega_r$.)
Here, we do not include charge noise, so the main features in the plot are caused by leakage.
However, a smaller contribution can also be ascribed to the counter-rotating terms. 
To demonstrate strong-driving effects more vividly, in Fig.~S1b we plot the fidelity of single-qubit gate operations at a TSS, as function of $\Delta B$ and $\tau$, while holding $\Delta_0$ fixed.
(Again, this requires varying $\omega_r$.
This result differs from Fig.~3d in the main text, where $\omega_r$ is held fixed.)
Since the $X_\pi$ gate time ($t=t_g$) depends only on $\varepsilon_\text{AC}d_{01}$, which is a non-oscillating, monotonic function of $\tau$ and $\Delta B$, the presence of oscillations in this figure can only be attributed to strong driving.

\section{Two-Qubit Gate Fidelity Oscillations}
The qubit decay rate is known to be enhanced when the qubit and cavity are resonant, due to the Purcell effect\citep{BlaisPRA2004}. 
Detuning the qubit(s) from the cavity suppresses this effect. 
In particular, in the dispersive regime, the Purcell decay is given asymptotically by $\gamma_P\approx (\hbar g/\Delta_0)^2\kappa$. 
In Fig.~4a of the main text, we observe this type of behavior near the resonance condition.
Further away from the resonance condition, the fidelity quickly reaches its maximum value and begins to decrease again.
In the main text, we explain this behavior as arising from the cumulative effects of dephasing, as the gate time becomes very long when the detuning $\Delta_0$ is large.
In addition to these general trends, Fig.~4a also exhibits small, fast oscillations very near the resonance condition.
In Fig.~S2, we show a blown-up view of those results, for the same device parameters. 
We now show that the oscillations are caused by a combination of leakage and strong driving.

The full Hamiltonian of the qubit-cavity-qubit system is given in Eq.~(5) of the main text as
\begin{equation}
H_{qqr}= \hbar\omega_r a^\dagger a +\sum_{i=a,b}\sum_n E_{n,i}\sigma_{nn,i} 
+\sum_{i=a,b}\sum_{n,m}\hbar g_{nm,i}(a+a^\dagger)\sigma_{nm,i},
\label{eq:HqqrSM}
\end{equation}
After eliminating the leakage states, as in the previous section, and moving to the rotating frame defined by 
\begin{equation}
H_0=\frac{\hbar\omega_{q,a}}{2}\sigma_{z,a}+\frac{\hbar\omega_{q,b}}{2}\sigma_{z,b}+\hbar\omega_ra^\dagger a,
\end{equation}
Eq.~(\ref{eq:HqqrSM}) reduces to
\begin{equation}
H_\text{int}=\sum_{j=a,b} \left[ \hbar g_j(a^\dagger\sigma_{-,j}e^{i\Delta_0 t/\hbar}+a\sigma_{+,j}e^{-i\Delta_0 t/\hbar})
+\hbar g_j(a^\dagger\sigma_{+,j}e^{i(\omega_{q,j}+\omega_r) t/\hbar}+a\sigma_{-,j}-e^{-i(\omega_{q,j}+\omega_r) t/\hbar})\right] .
\end{equation}
Here, the first term is responsible for the effective iSWAP interaction (after eliminating the photon states via a canonical transformation).
It can also cause leakage by exciting photons in the cavity, as well as the Purcell effect, when $\kappa >0$.
The second term contains counter-rotating (i.e., strong driving) terms and is responsible for the fast oscillations observed in Fig.~S2.
It is interesting to note that, for the case of a high quality factor (e.g., $Q=10^5$ used in this simulation), the effect of the counter-rotating terms can actually dominate over the Purcell decay, allowing the fidelity to recover its high value in very narrow bands, as shown in the figure.
However exploiting this effect would require exceptionally fine tuning of the device.

\begin{figure}[t]
\includegraphics[width=2.5 in]{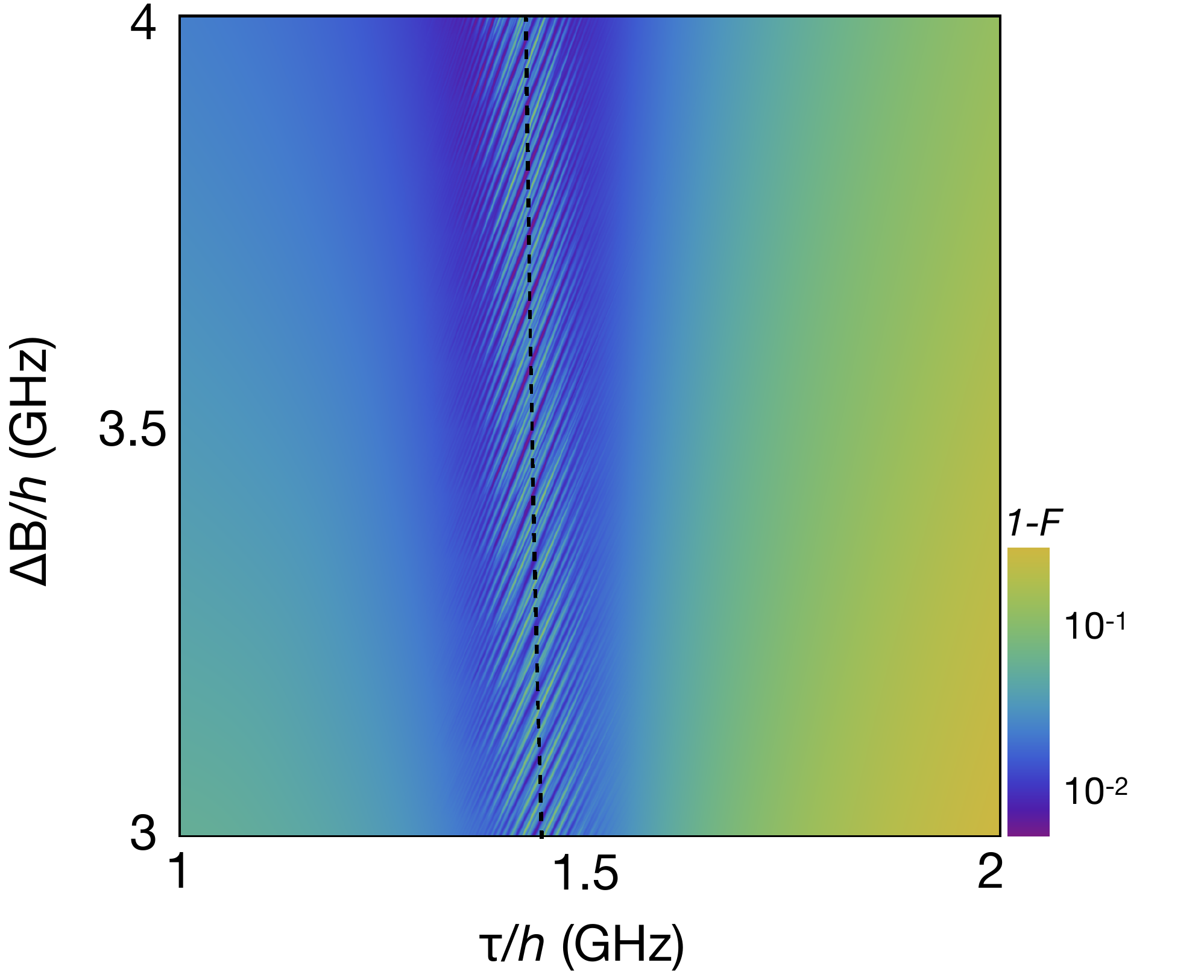}
\caption{A blown-up view of the results in Fig.~4a of the main text, using all the same simulation parameters.
The dashed black line marks the resonance condition.}
\label{figS2}
\end{figure} 

\section{Including a Global $B$-field and Transverse Field Gradients}
In Fig.~4b of the main text, we explore the two-qubit iSWAP gate errors caused by leakage into the spin-polarized states. In this section, we provide details of that calculation.

The two-qubit iSWAP gate fidelity is computed in two steps, following the same procedure used for simulations with just the spin-non-polarized states.
First, we expand the singlet-triplet Hamiltonian in Eq.~(1) of the main text to include the polarized triplet states, $\ket{T_-}=\ket{\uparrow\uparrow}$ and $\ket{T_+}=\ket{\downarrow\downarrow}$.

For the basis set  $\{|S(1,1)\rangle,|T_0(1,1)\rangle,|S(0,2)\rangle,|S(2,0)\rangle,\ket{T_-(1,1)},\ket{T_+(1,1)}\}$ we then obtain
\begin{equation}
H_\text{ST}=\begin{pmatrix}
0 & \Delta B & \sqrt{2}\tau & \sqrt{2}\tau & -\frac{\Delta B_\perp}{\sqrt{2}} & \frac{\Delta B_\perp}{\sqrt{2}}\\
\Delta B & 0 & 0 & 0 & 0 & 0 \\
\sqrt{2}\tau & 0 & -\varepsilon & 0 & 0 & 0 \\
\sqrt{2}\tau & 0 & 0 & 2U+\varepsilon & 0 & 0 \\
-\frac{\Delta B_\perp}{\sqrt{2}} & 0 & 0 & 0 & B & 0 \\
\frac{\Delta B_\perp}{\sqrt{2}} & 0 & 0 & 0 & 0 & -B
\end{pmatrix} .
\end{equation}
For simplicity, we may eliminate the states $\ket{S(2,0)}$ and $\ket{T_-(1,1)}$ from this equation, because they are very high in energy.
We then perform free-induction simulations in the presence of charge noise, with the initial state $\rho(t=0)=\ket{i}\bra{i}$, where $\ket{i}=(\ket{0}+\ket{1})/\sqrt{2}$ and $\ket{0}$ and $\ket{1}$ are the two low-energy qubit eigenstates. 
Averaging over noise realizations, as described in the main text, we obtain a final result for the density matrix $\rho(t)$.
Fitting this to a relaxation equation of form
\begin{equation}
1-|\rho_{T_+}(t)|
=\frac{1}{4}\exp\left[-(t/T_1) \right] + \frac{3}{4} 
\end{equation}
where $\rho_{T_+}(t)=\bra{T_+(1,1)}\rho(t)\ket{T_+(1,1)}$,
yields the relaxation time $T_1$ for leakage to the $\ket{T_+(1,1)}$ state.
In our simulations, we assume $\Delta B_\perp=\Delta B$, which gives the correct order of magnitude for $\Delta B_\perp$, although its exact value depends on the geometry of the nanomagnet \citep{YonedaAPE2015}. 

In the second step of the procedure, we construct an appropriate qubit-cavity-qubit Hamiltonian $H$, in analogy with the main text, and solve the expanded master equation
\begin{multline}
\dot{\rho}=-\frac{i}{\hbar}[H,\rho]+\frac{\kappa}{2}(2a\rho a^\dagger-a^\dagger a\rho-\rho a^\dagger a)\nonumber \\ 
+\sum_{j=a,b} \Big[\frac{1}{2T_\varphi}(\sigma_{z,j}\rho\sigma_{z,j}-\rho)
+\frac{1}{2T_L}(2\sigma_{L,	j}\rho\sigma_{L,j}^{\dagger}-\sigma_{L,j}^{\dagger}\sigma_{L,j}\rho-\rho\sigma_{L,j}^{\dagger}\sigma_{L,j}) 
+\frac{1}{2T_1}(2\sigma_{L',	j}\rho\sigma_{L',j}^{\dagger}-\sigma_{L',j}^{\dagger}\sigma_{L',j}\rho-\rho\sigma_{L',j}^{\dagger}\sigma_{L',j})\Big] ,
\end{multline}
which incorporates the expanded two-qubit basis set, and the leakage time $T_1$ obtained above.
Here, we have defined $\sigma_{L',i}=\sum_{n=0,1} \sigma_{nT_+,i}$.
\end{widetext}

\bibliography{singlettriplet}
\end{document}